\documentclass[a4paper]{jpconf}
\usepackage{graphicx}
\usepackage{amssymb,amsmath}
\begin{document}
\title{Deep learning for Directional Dark Matter search}

\author{Artem Golovatiuk$^{1,2}$, Giovanni De Lellis$^{1,2,3}$, Andrey Ustyuzhanin$^{4,5}$}

\address{$^1$Universit\`{a} degli Studi di Napoli Federico II, Corso Umberto I 40, 80138 Naples, Italy}
\address{$^2$INFN Section of Naples, via Cintia 21, 80126 Naples, Italy}
\address{$^3$CERN, Esplanade des Particules 1, 1211 Geneva 23, Switzerland}
\address{$^4$National Research University Higher School of Economics, 20 Myasnitskaya Ulitsa, Moscow 101000, Russia}
\address{$^5$National University of Science and Technology MISIS, Leninsky pr., 4, Moscow 119049 Russia}

\ead{artem.golovatiuk@cern.ch}

\begin{abstract}
We provide an algorithm for detection of possible dark matter particle interactions recorded within NEWSdm detector. The NEWSdm (Nuclear Emulsions for WIMP Search directional measure) is an underground Direct detection Dark Matter search experiment. The usage of recent developments in the nuclear emulsions allows probing new regions in the WIMP parameter space. The directional approach, which is the key feature of the NEWSdm experiment, gives the unique chance of overcoming the ”neutrino floor”. Deep Neural Networks were used for separation between potential DM signal and various classes of background. In this paper, we present the usage of deep 3D Convolutional Neural Networks to take into account the physical peculiarities of the datasets and report the achievement of the required $10^4$ background rejection power.
\end{abstract}

\section{Introduction}
    One of the large unsolved problems of modern physics is Dark Matter (DM). There are numerous observations both on astrophysical and cosmological length scales pointing to the existing of Dark Matter, but still no evidence in ground experiments. Three ways of looking for DM signatures on the experiment are: direct detection (detection of the DM-nuclei scattering), collider searches (creating of DM particles in the Standard Model collisions) and indirect detection (detection of the DM annihilation signals). The region in parameter space that can be tested by current direct detection experiments corresponds to the class of DM models called WIMPs (Weakly Interacting Massive Particles).

    \subsection{Directional detection}
    The directional dark matter search is a specific type of direct detection experiments. The main peculiarity of these experiments is the capability of measuring the direction of the signal. This gives a possibility to extend Dark Matter searches beyond the neutrino background and to prove the Galactic origin of the DM, since the flow of DM particles (so called 'WIMP wind') is expected to be directed fom the Cygnus constellation \cite{dm-direct} and the background is expected to be isotropic.
    
    \subsection{NEWSdm experiment}
    This study is carried out within the NEWSdm (Nuclear Emulsions for WIMP Search directional measure \cite{NEWSdm}) experiment. The experiment allows obtaining images of the DM-nuclei recoil tracks. The challenge is to recognise signal from dark-matter induced recoils from any other activity that may mimic it.
    
    The distinct features of the NEWS experiment among directional experiments are high sensitivity and possibility to scale the target mass due to usage of nuclear emulsions both as solid target and tracking device. To extract the information about the tracks below the resolution of the optical microscope the polarised light is used. The idea is based on the Plasmon resonance effect \cite{microscope} which gives reflected light the dependence on the crystal orientation, light polarisation angle and wavelength. This causes different behaviour of reflected light from the crystals within the same track.

\section{Experimental data}
    Training and test data is acquired by exposure of nuclear emulsions to a specific source, since Monte Carlo simulations are not available at the same level of details for this experiment. The exposed emulsions are chemically developed and scanned using an optical microscope with polarised light. Tracks from DM-nucleus interactions are simulated with Carbon ion beam with fixed energy (30-100 keV), corresponding to the WIMP mass range. The background is represented by several gamma radiation samples (Cs137 and Co60) which induce electrons in the emulsions and randomly excited grains by thermal fluctuations ("fog"). 
    
    Each sample consists of $\sim 10^5$ clusters, each of them having 8 monochromatic images for different light polarisation angles in the microscope during scanning. The exact number of clusters is presented in table~\ref{data_samples}. We use h5py package \cite{h5py} for Python \cite{python} to conveniently store our datasets in hdf5 format.

\begin{table}[h]
\caption{\label{data_samples}Number of clusters used during training and test for each emulsion sample.}
\begin{center}
\lineup
\begin{tabular}{*{3}{l}}
\br                              
Sample&Train&Test\cr 
\mr
Carbon/100keV&108604&72404\cr
Carbon/60keV&101536&67692\cr 
Carbon/30keV&104121&69414\cr 
fog&\056971&37981\cr 
gamma/Co60&\069870&46580\cr 
gamma/Cs137&\058900&39268\cr 
\br
\end{tabular}
\end{center}
\end{table}

\begin{figure}[h]
\begin{minipage}{17.5pc}
\includegraphics[width=16.5pc]{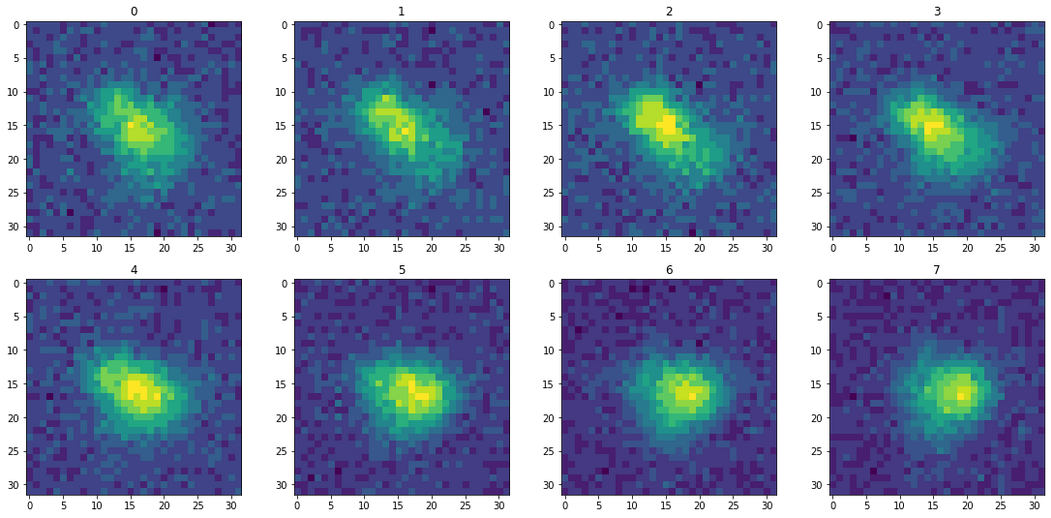}
\caption{\label{c100}8 polarisation images of a single Carbon 100keV track.}
\end{minipage}\hspace{2pc}%
\begin{minipage}{17.5pc}
\includegraphics[width=16.5pc]{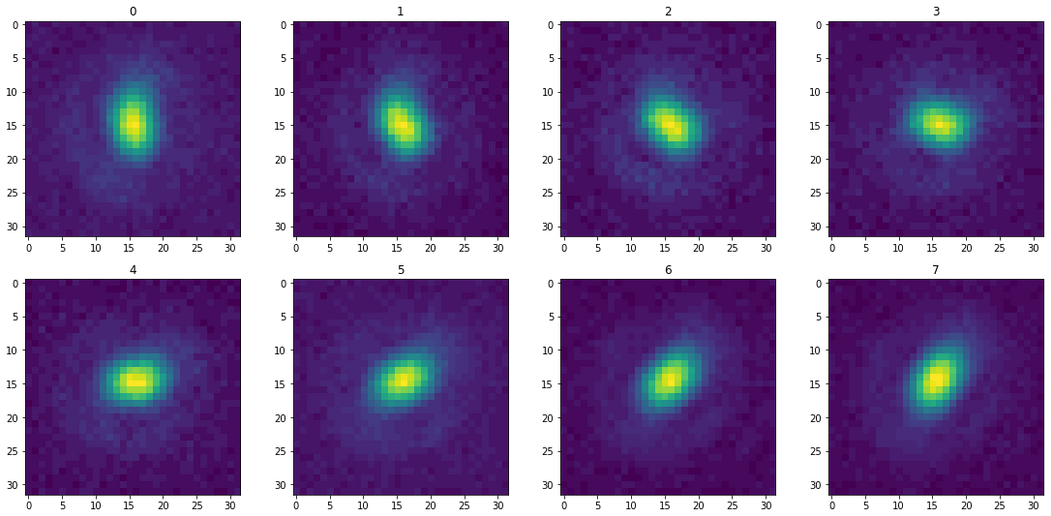}
\caption{\label{fog}8 polarisation images of a single fog background cluster.}
\end{minipage} 
\end{figure}

    Figures~\ref{c100} and~\ref{fog} explicitly demonstrates the Plasmon resonance effect on different cluster types. On figure~\ref{c100} the diagonal movement of the brightness peak can be seen, while on figure~\ref{fog} the peak is static, but the cluster itself is rotating. These are the distinctive features of signal tracks consisting of multiple grains and background single grains.
    
\section{Network and approach}
    This study is done using Python language \cite{python} and Keras package \cite{keras} for the Deep Learning algorithms. We use Convolutional Neural Networks (CNNs) for the signal-background classification. Our goal is the best background separation.
    
\begin{figure}
\centering
\includegraphics[width=32pc]{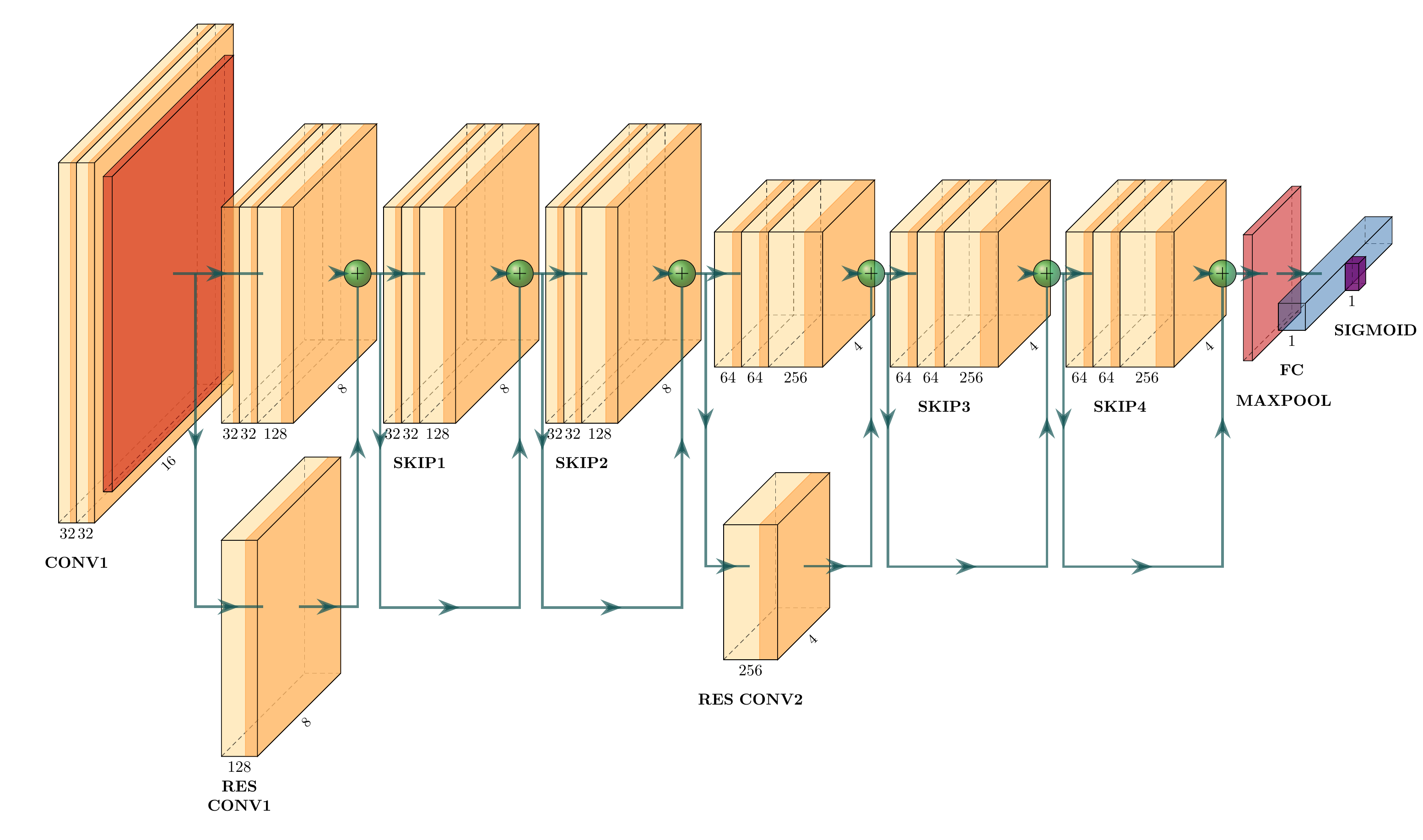}
\caption{\label{cnn}The schematic view of the residual CNN architecture.}
\end{figure}

    The network consists of basic building blocks like convolution layers, max-pooling layers and SWISH activation functions~\cite{swish}. The architecture is formed of convolutions and groups of residual skip connections \cite{resnet}. Filters within convolution layers are of the size $3\!\times\!3\!\times\!3$, while residual connections contain also layers with $1\!\times\!1\!\times\!1$ filters. Final output is obtained by applying a sigmoid activation function. Graphical representation of the architecture is shown in figure~\ref{cnn}.
    
    \subsection{Periodic boundary conditions}
    8 images for each cluster physically correspond to 8 discrete angular positions of the light polariser in the microscope and cover the whole $180^\circ$ rotation. Therefore, we add a copy of the 1st image after the 8th to imply periodic boundary conditions over the "polarisation axis". This allows network to see 8th and 1st images as neighbours and hence explore some correlations.
    The crystals in the track, that respond to polarised light, grow during chemical development in random directions, hence duplicating information in one polarisation angle image is not expected to bring any physically relevant bias.

    \subsection{3D CNN}
    As we already mentioned, Plasmon resonance effect~\cite{microscope} adds to the images important feature given by the "polarisation axis". To investigate correlations between images with different polariser angle we stack 9 images together to obtain a 3D image and use the 3D architecture of our CNN (the 3D versions of all components). Since we use monochromatic camera in the microscope, we do not have colour information in our images, so the input images are 4-dimensional tensors (2D image, polarisation axis and colour axis) with only one component in the colour axis.
    
    \subsection{Cluster rotations}
\begin{figure}
\centering
\includegraphics[width=34pc]{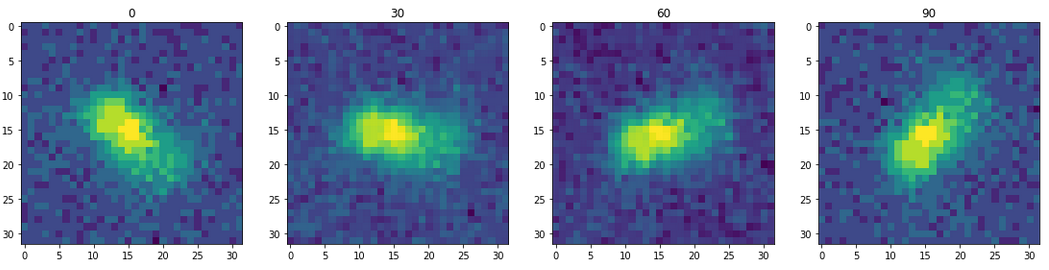}
\caption{\label{rotate}Example of the rotations of a single polarisation image of the Carbon track.}
\end{figure}

    The direction of the cluster is an important physical feature, that should be checked after selecting a signal candidate to verify its DM origin. Moreover, during the emulsion exposure the Carbon beam has a specific direction, while the background  is considered isotropic. Therefore, we apply random 2D rotations to each training batch (256 clusters) to make the training data isotropic and ensure that the network is not using directionality as a distinctive feature for its classification. An example of a rotated cluster image is shown in figure~\ref{rotate}. All 9 polarisation images of a cluster are rotated together. We compare the performances of the networks trained with and without cluster rotations.
    
\section{Background rejection performance}
\begin{figure}[h]
\includegraphics[width=22pc]{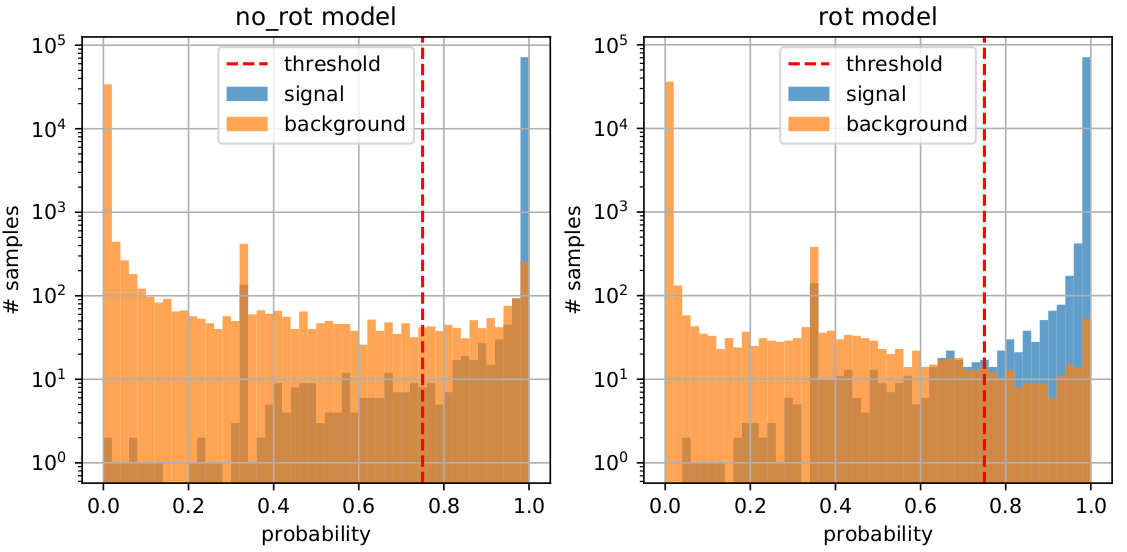}\hspace{2pc}%
\begin{minipage}[b]{13pc}\caption{\label{c100_pro}Network output on a test sample for Carbon 100keV against fog. Red line demonstrates one possible classifier output threshold.}
\end{minipage}
\end{figure}

    The default output of the algorithm for a cluster is a probability of being signal. To construct an algorithm with binary output and compute the metrics we need to specify a threshold on the probability output. The metrics we use are background rejection power - all input "background" divided by false claimed "signal" (remaining "background" contamination), and efficiency - true "signal" divided by all input "signal". Figure~\ref{c100_pro} compares the output histograms of the two training approaches (with and without random rotations). The scale of the Y-axis is logarithmic.
    
\begin{figure}
\centering
\includegraphics[width=34pc]{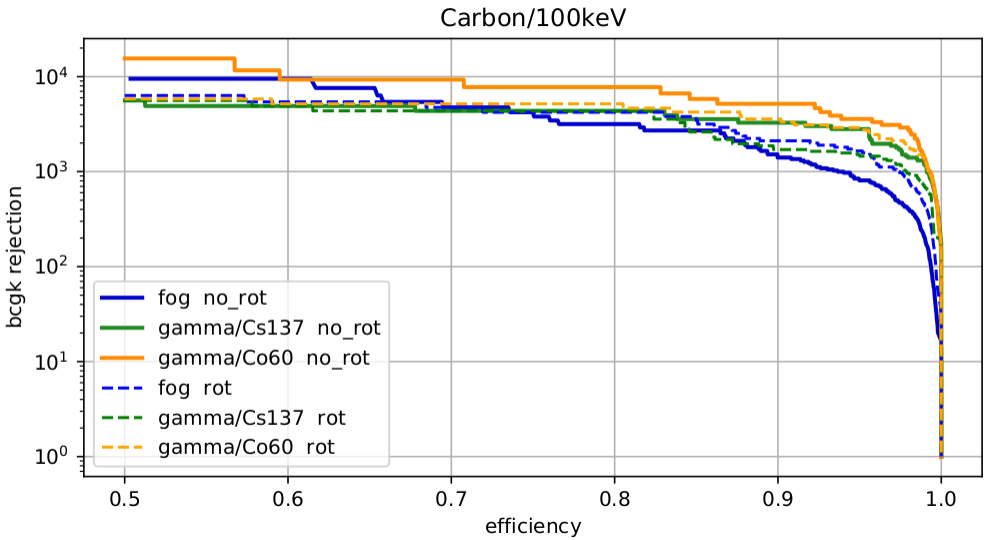}
\caption{\label{c100_rej}Background rejection - efficiency plot for Carbon 100 keV against background. Different points on the curve correspond to different threshold values for the probabilistic output.}
\end{figure}

    Current background rejection power goal for the NEWSdm experiment is $\sim 10^4$ with the efficiency of $\gtrsim 0.6$. The final exact background rejection goal will depend on the background source and a combination of other methods of rejection. Here on figure~\ref{c100_rej} we compare the test performances of the models trained with and without random image rotations in classification task for Carbon 100keV ion tracks against various background types.
    
    One can see that the current goal is already achievable for both "fog" and gamma (Co60) background types, but only for the CNN trained without random image rotations. However, eliminating the directionality feature of the clusters gives important benefits for further physical analysis of the sample, so "rotational" models require some additional improvements.\\
    Different behaviour of gamma from Cs137 and Co60 can be caused by the different decay energies or by some peculiarities of the emulsion samples.
    
\section{Summary and conclusions}
    We studied signal-background classification with CNNs in NEWSdm - the nuclear emulsion Dark Matter search experiment. Analysis of the physical symmetries and peculiarities of the detector reflected both by the network’s architecture and data preprocessing pipeline boosted the classification performance. The algorithm performance achieves the experimentally required background rejection power of  $10^4$ for two out of three kinds of background. 
    
\ack
    We are grateful to all the members of the Naples group for assistance in the emulsion scanning process, all the discussions of new ideas and approaches.\\
    The research leading to these results has received funding from Russian Science Foundation under grant agreement n° 17-72-20127.
    This work is supported by a Marie Sklodowska-Curie Innovative Training Network Fellowship of the European Commissions Horizon 2020 Programme under contract number 765710 INSIGHTS.

\end{document}